\documentstyle[12pt]{article}
\evensidemargin\oddsidemargin
 \catcode`@=11
\def\@cite#1#2{\unskip\nobreak\relax
    \def\@tempa{$\m@th^{\hbox{\the\scriptfont0 #1}}$}%
    \futurelet\@tempc\@citexx}
\def\@citexx{\ifx.\@tempc\let\@tempd=\@citepunct\else
    \ifx,\@tempc\let\@tempd=\@citepunct\else
    \let\@tempd=\@tempa\fi\fi\@tempd}
\def\@citepunct{\@tempc\edef\@sf{\spacefactor=\the\spacefactor\relax}\@tempa
    \@sf\@gobble}

\def\citenum#1{{\def\@cite##1##2{##1}\cite{#1}}}
\def\citea#1{\@cite{#1}{}}

\newcount\@tempcntc
\def\@citex[#1]#2{\if@filesw\immediate\write\@auxout{\string\citation{#2}}\fi
  \@tempcnta\z@\@tempcntb\m@ne\def\@citea{}\@cite{\@for\@citeb:=#2\do
    {\@ifundefined
       {b@\@citeb}{\@citeo\@tempcntb\m@ne\@citea\def\@citea{,}{\bf ?}\@warning
       {Citation `\@citeb' on page \thepage \space undefined}}%
    {\setbox\z@\hbox{\global\@tempcntc0\csname b@\@citeb\endcsname\relax}%
     \ifnum\@tempcntc=\z@ \@citeo\@tempcntb\m@ne
       \@citea\def\@citea{,}\hbox{\csname b@\@citeb\endcsname}%
     \else
      \advance\@tempcntb\@ne
      \ifnum\@tempcntb=\@tempcntc
      \else\advance\@tempcntb\m@ne\@citeo
      \@tempcnta\@tempcntc\@tempcntb\@tempcntc\fi\fi}}\@citeo}{#1}}
\def\@citeo{\ifnum\@tempcnta>\@tempcntb\else\@citea\def\@citea{,}%
  \ifnum\@tempcnta=\@tempcntb\the\@tempcnta\else
   {\advance\@tempcnta\@ne\ifnum\@tempcnta=\@tempcntb \else \def\@citea{--}\fi
    \advance\@tempcnta\m@ne\the\@tempcnta\@citea\the\@tempcntb}\fi\fi}
\catcode`@=12

\renewenvironment{thebibliography}[1]
 {\begin{list}{\arabic{enumi}.}
    {\usecounter{enumi} \setlength{\parsep}{0pt}
     \setlength{\itemsep}{3pt} \settowidth{\labelwidth}{#1.}
     \sloppy
    }}{\end{list}}

\newcommand{\alt}{\mathrel{\raisebox{-.6ex}{$\stackrel{\textstyle<}{\sim}$}}}
\newcommand{\agt}{\mathrel{\raisebox{-.6ex}{$\stackrel{\textstyle>}{\sim}$}}}

\def\overlay#1#2{\ifmmode \setbox 0=\hbox {$#1$}\setbox 1=\hbox to\wd 0{\hss
$#2$\hss }\else \setbox 0=\hbox {#1}\setbox 1=\hbox to\wd 0{\hss #2\hss }\fi
#1\hskip -\wd 0\box 1}

\def\nv#1 {\noalign{\vskip#1pt}}

\def\abstract#1{\begin{center}\Large\bf Abstract\end{center}
{\narrower\small #1\par}}

\def\gev{{\rm\,GeV}}

\def\L{{\cal L}}
\def\O{{\cal O}}

\def\fb{{\rm\,pb}}

\begin{document}

\font\fortssbx=cmssbx10 scaled \magstep2
\hbox to \hsize{
\hskip.5in \raise.1in\hbox{\fortssbx University of Wisconsin - Madison}
\hfill\vbox{\hbox{\bf MAD/PH/762}
            \hbox{May 1993}} }

\vspace{.25in}

\begin{center}
{\LARGE\bf CLOSING IN ON SUPERSYMMETRY\footnote{Talk presented by V.~Barger at
the {\it Symposium in Honor of Tetsuro Kobayashi's 63rd Birthday}, Tokyo, March
1993}}\\[.4in]
{\large V.~Barger$^{\,a}$  and R.J.N.~Phillips${\,^b}$}\\[.2in]
\it
$^a$Physics Department, University of Wisconsin, Madison, WI 53706, USA\\
$^b$Rutherford Appleton Laboratory, Chilton, Didcot, Oxon OX11 0QX, UK
\end{center}

\renewcommand{\LARGE}{\Large}
\renewcommand{\Huge}{\Large}

\vspace{.25in}

\abstract{
A survey is made of some recent ideas and progress in the
phenomenological applications of Supersymmetry (SUSY).
We describe the success of SUSY-GUT models, the expected
experimental signatures and present limits on SUSY partner particles,
and the phenomenology of Higgs bosons in the minimal SUSY model.}

\vspace{.5in}

\section{Introduction}

   The Standard Model (SM) has been very succesful so far;
all its predictions that have been tested have been
verified to high precision.  Important checks remain to be made, of course:
the top quark is not yet discovered, the interactions between gauge bosons
are still unmeasured, and the Higgs boson remains a totally unconfirmed
hypothesis.  In spite of its success, however, the apparent
arbitrariness and various theoretical limitations of the SM suggest
a need for a deeper theory such as Supersymmetry
(SUSY) or Superstrings, implying new physics, new particles
and new interactions beyond the SM.  In this lecture we review some recent
developments in SUSY phenomenology: unification of couplings in
SUSY-GUT models, experimental signals from SUSY, and Higgs phenomenology
in the minimal SUSY extension of the SM (MSSM).

   With SUSY\cite{reviews}, each fermion has a boson partner (and vice versa),
with all the same quantum numbers but with spin differing by 1/2.  Since
no such partners have been found, SUSY is plainly a broken symmetry at
presently accessible mass scales but could be restored above some higher scale
$M_{\rm SUSY}$.

   The primary theoretical motivation for SUSY is that it stabilizes divergent
loop contributions to scalar masses, because fermion and boson loops
contribute with opposite signs and largely cancel.  This cures the
naturalness problem in the SM, so long as $M_{\rm SUSY} \alt \O(1\rm\ TeV)$,
where otherwise the Higgs mass would
require fine-tuning of parameters. There are also attractive practical
features:  SUSY-GUT models can be
calculated perturbatively and can be tested experimentally at supercolliders,
where SUSY partners can be produced and studied.  Philosophically, SUSY
is the last possible symmetry of the $S$-matrix~\cite{haag},
and there is a predisposition to believe that anything not forbidden is
compulsory.

   Phenomenological interest has focussed mainly on the  MSSM,
which introduces just one
spartner for each SM particle.  The gauge symmetry is $\rm SU(3)_c\times
SU(2)_L\times U(1)_Y$; the corresponding spin-1 gauge bosons $g,W,Z,\gamma$
have spin-1/2 ``gaugino" partners $\tilde g,\tilde W, \tilde Z, \tilde \gamma$.
The three generations of spin-1/2 quarks $q$ and leptons $\ell$ have spin-0
squark and slepton partners $\tilde q$ and $\tilde\ell$; the
chiral states $f_L$ and $f_R$ of any given fermion $f$ have distinct sfermion
partners $\tilde f_L$ and $\tilde f_R$, respectively (that can however mix).
For anomaly cancellation the single Higgs doublet must be replaced by two
doublets $H_1$
and $H_2$ that have higgsino partners $\tilde H_1$ and $\tilde H_2$.
The MSSM also conserves a multiplicative $R$-parity, defined by
\begin{equation}
    R = (-1)^{2S+L+3B}
\end{equation}
where $S,L,B$ are spin, lepton number and baryon number.  The normal particles
of the SM all have $R=+1$;  their spartners
which differ simply by 1/2 unit of $S$, therefore have $R=-1$.
$R$-conservation  has important physical implications:

\begin{enumerate}
\item sparticles must be produced in pairs,
\item heavy sparticles decay to lighter sparticles,
\item the lightest sparticle (LSP) is stable.
\end{enumerate}
If this LSP has zero charge and only interacts weakly, as seems likely
since it has defied detection so far, it will carry off undetected
energy and momentum in high-energy collisions (providing possible
signatures for sparticle production) and will offer a possible source of
cosmological dark matter.

In addition to the more general motivations
above,  there are also several significant phenomenological motivations
for SUSY.

\begin{enumerate}

\item Grand Unified Theories (GUTs) with purely SM particle content do not
predict a satisfactory convergence of the gauge couplings at some high
GUT scale $M_G$, but convergence can be achieved if SUSY partners are added
(see Section~2)~\cite{amaldi,ellis,langacker}.

\item Starting from equal $b$ and $\tau$ Yukawa couplings at the GUT scale
$M_G$, the physical masses can be correctly predicted when the evolution
equations include SUSY partners, but not with the SM alone (see
Section~2)~\cite{ramond,giveon}.

\item Proton decay is too rapid in a SM GUT but can be acceptable in
SUSY-GUT models where $M_G$ is higher~\cite{hisano}.

\item With $R$-parity conservation, the lightest SUSY partner (LSP) is
stable and provides a plausible candidate for the origin of dark matter
making $\Omega \sim 1$~\cite{drees,kelley,roberts}.

\item SUSY-GUT models naturally cause the Higgs field to develop a vacuum
expectation value, when the top mass is larger than $M_W$~\cite{ross}.
\end{enumerate}

\section{SUSY-GUT models}

    As the renormalization mass scale $\mu$
is changed, the evolution of couplings is governed by the
Renormalization Group Equations (RGE).
For the gauge group $\rm SU(3)\times SU(2)\times U(1)$,
with corresponding gauge couplings
$g_3(=g_s), g_2(=g), g_1(=\sqrt{5/3}g')$, the RGE can be written in
terms of the dimensionless variable $t=\ln(\mu/M_G)$:
\begin{equation}
{dg_i\over dt} = {g_i\over 16\pi^2} \left[ b_ig_i^2 + {1\over16\pi^2}
\left( \sum_{j=1}^3 b_{ij} g_i^2g_j^2 - \sum a_{ij}
g_i^2\lambda_j^2\right)\right]
\end{equation}
The first term on the right is the one-loop approximation; the second
and third terms contain two-loop effects, involving other gauge
couplings $g_j$ and Yukawa couplings $\lambda_j$.  The coefficients $b_i,\
b_{ij}$ and $a_{ij}$ are determined at given scale $\mu$ by the content of
active particles  (those with mass ${}<\mu$).  If there are no thresholds
({\it i.e.} no
changes of particle content) between $\mu$ and $M_G$, then the coefficients
are constants through this range and the one-loop solution is
\begin{equation}
\alpha_i^{-1}(\mu)  =  \alpha_i^{-1}(M_G) - t b_i/(2\pi)   \;,
\end{equation}
where $\alpha_i = g_i^2/(4\pi)$; thus $\alpha_i^{-1}$ evolves linearly with
$\ln\mu$ at one-loop order.    If there are no new physics thresholds
between $\mu = M_Z \simeq m_t$  and $M_G$  ({\it i.e.} just a ``desert" as in
the basic SM) then equations of this kind should evolve the observed
couplings at the electroweak scale~\cite{giatw}
\begin{eqnarray}
\alpha_1(M_Z)^{-1} &=& 58.89 \pm 0.11 \,, \\
\alpha_2(M_Z)^{-1} &=& 29.75 \pm 0.11 \,, \\
\alpha_3(M_Z)^{-1} &=& 0.118 \pm 0.007 \,,
\end{eqnarray}
to converge to a common value at some large scale.  Figure~1(a) shows that
such a SM extrapolation does NOT converge; this figure actually includes
two-loop effects but the evolution is still approximately linear versus
$\ln\mu$, as at one-loop order.  GUTs do not work, if we assume just SM
particles plus a desert up to $M_G$.

   But if we increase the particle content to include the minimum
number of SUSY particles, with a threshold not too far above $M_Z$,  then
GUT-type convergence can happen.  Figure~1(b) shows an example with SUSY
threshold $M_{\rm SUSY}=1$~TeV~\cite{bbo}.  SUSY-GUTs are plainly
more successful; the evolved couplings are consistent with a common
intersection at $M_G \sim 10^{16}$\,GeV.  In fact a precise single-point
intersection is not strictly necessary, since the exotic GUT gauge,
fermion and scalar particles do not have to be precisely degenerate;
we may therefore have several non-degenerate thresholds near $M_G$, to be
passed through on the way to GUT unification.

\begin{center}
\parbox{5.5in}{\small Fig.~1. Gauge coupling evolution: (a)  in the SM;
(b)  in a SUSY-GUT example~\cite{bbo}.}
\end{center}

   The Yukawa couplings also evolve.  The evolution equations for $\lambda_t$
and $\lambda_b/\lambda_\tau$ are
\begin{equation}
{d\lambda_t\over dt} = {\lambda_t\over16\pi^2} \left[-\sum c_i g_i^2 +
6\lambda_t^2 + \lambda_b^2  + \mbox{2-loop terms}\right] \;, \label{yuklam_t}
\end{equation}
with $c_1=13/15$, $c_2=3$, $c_3=16/3$, and
\begin{equation}
{d(\lambda_b/\lambda_\tau)\over dt} = {(\lambda_b/\lambda_\tau)\over16\pi^2}
\left[-\sum d_i g_i^2+\lambda_t^2+3\lambda_b^2-3\lambda_\tau^2
+ \mbox{2-loop terms} \right]\,, \label{yuklam_b}
\end{equation}
with $d_1=-4/3$, $d_2=0$, $d_3=16/3$. The low-energy values at $\mu=m_t$ are
\begin{eqnarray}\label{lambda_b}
\lambda_b(m_t) = {\sqrt2\, m_b(m_b)\over\eta_b v\cos\beta}\,, \qquad
\lambda_\tau(m_t) = {\sqrt2m_\tau(m_\tau)\over \eta_\tau v\cos\beta}\,, \qquad
\lambda_t(m_t) = {\sqrt2 m_t(m_t)\over v\sin\beta} \;,
\end{eqnarray}
where $\eta_f = m_f(m_f)/m_f(m_t)$ gives the running of the masses below
$\mu=m_t$, obtained from 3-loop QCD and 1-loop QED evolution. The $\eta_q$
values depend on the value of $\alpha_3(M_Z)$; for $\alpha_3(M_Z)=0.118,\
\eta_b=1.5,\ \eta_c=2.1,\ \eta_s=2.4$. The running mass values are
$m_\tau(m_\tau)=1.777$~GeV and $m_b(m_b)=4.25\pm0.15$~GeV.  The denominator
factors in Eq.~(\ref{lambda_b}) arise from the two Higgs vevs $v_1=v\cos\beta$
and $v_2=v\sin\beta$; they are related to the SM vev $v=246$~GeV by
$v_1^2+v_2^2=v^2$, while $\tan\beta=v_2/v_1$ measures their ratio.

It is frequently  assumed that the $b$-quark and
$\tau$-lepton Yukawa couplings are equal at the GUT scale:~\cite{eight,dhr}
\begin{equation}
\lambda_b(M_G) = \lambda_\tau(M_G) \;. \label{b=tau}
\end{equation}
Figure 2 illustrates the running of $\lambda_t$, $\lambda_b$ and
$\lambda_\tau$, obtained from solutions to the RGEs with the appropriate
low-energy boundary conditions and the GUT-scale condition of
Eq.~(\ref{b=tau}). Note that  $\lambda_t(M_G)$ must be large in order to
satisfy the  boundary condition
$m_b(m_b)=4.25\pm0.15$.

\begin{center}
{\small Fig.~2. The running of $\lambda_t$, $\lambda_b$ and $\lambda_\tau$ from
low energies to the GUT scale~\cite{bbo}.}
\end{center}

As $\mu\to m_t$, $\lambda_t$ rapidly approaches a fixed point~\cite{pendleton}.
The approximate fixed-point solution for $m_t$ is
\begin{equation}
-\sum c_ig_i^2 + 6\lambda_t^2 + \lambda_b^2 = 0 \,.
\end{equation}
Neglecting $g_1,\, g_2$ and $\lambda_b$, $m_t$ is predicted in terms of
$\alpha_s(m_t)$ and $\beta$:~\cite{dhr,bbhz}
\begin{eqnarray}
m_t(m_t) \approx \frac{4}{3}\sqrt{2\pi\alpha_s(m_t)}\,
\frac{v}{\sqrt2}\sin\beta
\approx (200\gev){\tan\beta\over\sqrt{1+\tan^2\beta}} \,. \label{mt(mt)}
\end{eqnarray}
Thus the natural scale of the top-quark mass is large in SUSY-GUT models. Note
that the propagator-pole mass is related to this running mass by
\begin{equation}
m_t({\rm pole}) = m_t(m_t)\left[1+{4\over3\pi}\alpha_s(m_t)+\cdots\right]\,.
\label{mtpole}
\end{equation}

An exact numerical solution for the relation between $m_t$ and $\tan\beta$,
obtained from the 2-loop RGEs for $\lambda_t$ and $\lambda_b/\lambda_\tau$, is
shown in Fig.~3 taking $M_{\rm SUSY}=m_t$. At large $\tan\beta$, $\lambda_b$
becomes large and the above fixed-point solution no longer applies. In fact,
the solutions becomes non-perturbative at large $\tan\beta$ and we impose the
perturbative requirements $\lambda_t(M_G)\leq3.3$, $\lambda_b(M_G)\leq3.1$,
based on the requirement that (2-loop)/(1-loop)$\leq 1/4$ giving $\tan\beta\alt
65$. At large $\tan\beta$ there is the possibility of
$\lambda_t=\lambda_b=\lambda_\tau$ unification.
For most $m_t$ values there are two possible solutions for $\tan\beta$; the
lower solution is controlled by the $\lambda_t$ fixed point, following
Eqs.~(\ref{mt(mt)}),(\ref{mtpole}):
\begin{equation}
\sin\beta\simeq m_t(\rm pole)/(200\gev) \,.
\end{equation}
An upper limit $m_t(\rm pole) \alt200$~GeV is found with the RGE solutions.

\bigskip

\begin{center}
{\small Fig.~3. Contours of constant $m_b(m_b)$ in the
$\left(m_t(m_t),\tan\beta^{\vphantom1}\right)$ plane~\cite{bbo}.}
\end{center}

\bigskip

Figure 4 shows the dependence of $\lambda_t(M_G)$ on $\alpha_3(M_Z)$. For
$\lambda_t$ to remain perturbative, an upper limit $\alpha_3(M_Z)\alt0.125$ is
necessary.

\begin{center}
{\small Fig.~4. Dependence of  $\lambda_t$ at the GUT scale on
$\alpha_3(M_Z)$~\cite{bbo}.}
\end{center}

Specific GUT models also make predictions for CKM matrix elements. For example,
several models~\cite{dhr,hrr} give the GUT-scale relation
\begin{equation}
 |V_{cb}(\rm GUT)|=\sqrt{\lambda_c({\rm GUT})/\lambda_t({\rm GUT})}\,.
\end{equation}
The relevant RGEs are
\begin{eqnarray}
{d|V_{cb}|\over dt} &=&
-{|V_{cb}|\over16\pi^2}\left[\lambda_t^2+\lambda_b^2+\mbox{2-loop}\right]\,,\\
{d(\lambda_c/\lambda_t)\over dt} &=& -{(\lambda_c/\lambda_t)\over16\pi^2}
\left[3\lambda_t^2+\lambda_b^2+\mbox{2-loop}\right]\,,
\end{eqnarray}
\noindent
in addition to Eqs.~(\ref{yuklam_t}) and (\ref{yuklam_b}).
Starting from boundary conditions on $m_c$ and $|V_{cb}|$ at scale $\mu=m_t$,
the equations can be integrated up to $M_G$ and checked to see if the above
GUT-scale constraint is satisfied. The low-energy boundary conditions are
\begin{equation}
0.032 \le |V_{cb}(m_t)|\le0.054\,, \qquad 1.19\le m_c(m_c)\le1.35\gev \,.
\end{equation}

The resulting $|V_{cb}|$ solutions at the 2-loop level are shown  in Fig.~5.
The contours of $m_b(m_b)=4.1$ and 4.4~GeV, which satisfy
$\lambda_b(M_G)/\lambda_\tau(M_G)=1$, are also shown in Fig.~5(a). The shaded
region in Fig.~5(a) denotes the solutions that satisfy both sets of GUT-scale
constraints. A lower limit $m_t\ge155$~GeV can be inferred, based on values
$m_c=1.19$ and $\alpha_3(M_Z)=0.110$ used in this illustration; with
$\alpha_3(M_Z)=0.118$ instead, $m_t$ can be as low as 120~GeV with
$|V_{cb}|=0.054$. One GUT ``texture'' that leads to the above $|V_{cb}|$ GUT
prediction is given by the following up-quark, down-quark and lepton
mass-matrix structure at $M_G$~\cite{dhr}:
\begin{equation}
{\bf U}= \left(\begin{array}{ccc}
0&C&0\\ C&0&B\\ 0&B&A\end{array}\right) \qquad
{\bf D}= \left(\begin{array}{ccc}
0&Fe^{i\phi}&0\\ Fe^{-i\phi}&E&0\\ 0&0&D\end{array}\right) \qquad
{\bf E}= \left(\begin{array}{ccc}
0&F&0\\ F&E&0\\ 0&0&D\end{array}\right) .
\end{equation}

\begin{center}
\parbox{6in}{\small Fig.~5. Contours of constant $m_b(m_b)$ for
$\lambda_b/\lambda_\tau=1$ at $\mu=M_G$ and contours of constant
$|V_{cb}(m_t)|$, (a)~in the $\left(m_t(m_t),\sin\beta^{\phantom1}\right)$ plane
and (b)~in the $\left(m_t(m_t),\tan\beta^{\phantom1}\right)$
plane~\cite{bbo,bbhz}.}
\end{center}

\section{Experimental Signatures for SUSY}

Experimental evidence for SUSY could come in various forms, for example
\begin{enumerate}

\item discovery of one or more superpartners,

\item discovery of a light neutral Higgs boson with non-SM properties and/or a
charged Higgs boson,

\item discovery of $p\to K\nu$ decay: the present lifetime limit is $10^{32}$
years but Super-Kamiokande will be sensitive up to 10$^{34}$ years,

\item discovery that dark matter is made of heavy ($\alt 100$~GeV) neutral
particles.
\end{enumerate}

GUTs are essential for SUSY phenomenology; without GUTs there would be far too
many free parameters. A minimal set of GUT parameters with soft SUSY breaking
consists of the gauge and Yukawa couplings $g_i$ and $\lambda_i$, the Higgs
mixing mass $\mu$, the common gaugino mass at the GUT scale $m_{1/2}$, the
common scalar mass at the GUT scale $m_0$, and two parameters $A,B$ that give
trilinear and bilinear scalar couplings. At the weak scale, the gauge couplings
are experimentally determined. The Higgs potential depends upon $m_0,\mu,B$ (at
tree level) and $m_{1/2},A,\lambda_t,\lambda_b$ (at one loop). After minimizing
the Higgs potential and putting in the measured $Z$ and fermion masses, there
remain 5 independent parameters that can be taken as
$m_t,\tan\beta,m_0,m_{1/2},A$, though other independent parameter sets are
often used for specific purposes.

The SUSY particle spectrum consists of Higgs bosons $(h,H,A,H^\pm)$, gluinos
$(\tilde g)$, squarks $(\tilde q)$, sleptons $(\tilde\ell^\pm)$, charginos
($\tilde W_i^\pm, i=1,2$; mixtures of winos and charged higgsinos), neutralinos
($\tilde Z_j, j=1,2,3,4$; mixtures of zinos, photinos and neutral higgsinos).
An alternate notation is $\tilde\chi_i^+$ for $\tilde W_i^+$ and
$\tilde\chi_j^0$ for $\tilde Z_j$. The evolution of the SUSY mass spectrum from
the GUT scale~\cite{ross,rosrob} is illustrated in Fig.~6. The running masses
are plotted versus $\mu$ and the physical value occurs approximately where the
running mass $m=m(\mu)$ intersects the curve $m=\mu$. In the case of the Higgs
scalar $H_2$, the mass-square becomes negative at low $\mu$ due to coupling to
top; in this region we have actually plotted $-|m(\mu)|$. Negative mass-square
parameter is essential for spontaneous symmetry-breaking, so this feature of
SUSY-GUTs is desirable; it is achieved by radiative effects. The running masses
for the gauginos $\tilde g,\tilde W,\tilde B$ are given by
\begin{equation}
M_i(\mu)=m_{1/2}\,{\alpha_i(\mu)\over\alpha_i(M_G)} \,,
\end{equation}
where $i$ labels the corresponding gauge symmetry; this applies before we add
mixing with higgsinos to obtain the chargino and neutralino mass eigenstates.
In the example of Fig.~6 the squarks are heavier than the gluinos, but the
opposite ordering $m_{\tilde q}< m_{\tilde g}$ is possible in other scenarios.
Sleptons, neutralinos and charginos are lighter than both squarks and gluinos
in general.
Note that the usual soft SUSY-breaking mechanisms preserve the
gauge coupling relations (unification) at $M_G$.

   In order that SUSY cancellations shall take effect at low mass
scales as required, the SUSY mass parameters are expected to be
bounded by
\begin{equation}
      m_{\tilde g},\, m_{\tilde q},\, |\mu|,\, m_A \alt 1\mbox{--2 TeV}\, .
\label{m bound}
\end{equation}
The other parameter $\tan\beta$ is effectively bounded by
\begin{equation}
          1  \alt    \tan\beta  \alt    65    \;, \label{eff bound}
\end{equation}
where the lower bound arises from consistency in GUT models and the
   upper bound is the perturbative limit. Proton decay gives the constraint
$\tan\beta<85$~\cite{hisano}.

\begin{center}
{\small Fig.~6. Representative RGE results for spartner masses~\cite{ross}.}
\end{center}

      At LEP\,I, sufficiently light SUSY particles would be produced
through their gauge couplings to the $Z$.  Direct searches for
SUSY particles at LEP give mass lower bounds
\begin{equation}
      m_{\tilde q},\, m_{\tilde \ell},\,m_{\tilde\nu},\,
 m_{\tilde W_1}\agt\mbox{40--45 GeV}\;.
\end{equation}
The limitation of LEP is its relatively low CM energy.

      Hadron colliders can explore much higher energy ranges.
For $m_{\tilde q}=m_{\tilde q}$,  about 100 events would be expected
for each of the channels $\tilde g\tilde q$ and $\tilde q\tilde q$
at mass 200~GeV~\cite{bt,btw}, so the Tevatron clearly reaches well beyond the
LEP range.

      The most distinctive SUSY signature is the missing
energy and momentum carried off by the LSP, usually
assumed to be the lightest neutralino $\tilde Z_1$, which occurs in
all SUSY decay chains with $R$-parity conservation.  At hadron colliders
it is only possible to do book-keeping on the missing transverse momentum
denoted $\overlay/p_T$.   The missing momenta of both LSPs are added
vectorially in $\overlay/p_T$.  The LSP momenta and hence the magnitude
of  $\overlay/p_T$ depend on the decay chains.

      If squarks and gluinos are rather light ($m_{\tilde g},m_{\tilde q}\alt
50$~GeV), their dominant decay
mechanisms are direct strong decays or decays to the LSP:
\begin{eqnarray}
\left.\begin{array}{l}
      \tilde q \to q \tilde g\\
\tilde g \to q \bar q \tilde Z_1
      \end{array} \right\}
\quad{\rm if}\ m_{\tilde g} < m_{\tilde q} \,, \label{light a}\qquad
\left.\begin{array}{l}
           \tilde g\to q \tilde q \phantom{Z_1} \\
\tilde q\to  q \tilde Z_1  \phantom{q}
\end{array}\right\}
\quad {\rm if}\ m_{\tilde q} < m_{\tilde g} \,. \label{light b}
\end{eqnarray}
In such cases the LSPs carry a substantial fraction of the available
energy and $\overlay/p_T$ is correspondingly large.  Assuming such decays and
small LSP mass, the present 90\%~CL experimental bounds from UA1 and UA2
at the CERN $p$-$\bar p$ collider ($\sqrt s = 640$~GeV) and from CDF at the
Tevatron ($\sqrt s = 1.8$~TeV) are~\cite{uacdf}
\[
\vbox{\tabskip2em\halign{#\hfil&&$#$\hfil\cr
                    &  \hfil  m_{\tilde g}  &  \hfil  m_{\tilde q}\cr
        UA1 (1987)      &     >  53\rm\ GeV  &      >  45\rm\ GeV \cr
        UA2 (1990)      &        >  79       &     >  74 \cr
        CDF (1992)      &        > 141       &     > 126 \cr}}
\]
The limits become more stringent if $m_{\tilde g}$ and $m_{\tilde q}$
are assumed to be comparable.

   For heavier gluinos and squarks, many new decay channels are open,
such as
\begin{eqnarray}
  \tilde g  &\to& q \bar q \tilde Z_i\ (i=1,2,3,4),\ q \bar q' \tilde W_j\
                 (j=1,2),\ g \tilde Z_1 \;, \label{heavy a}\\
  \tilde q_L &\to& q \tilde Z_i\ (i=1,2,3,4),\ q' \tilde Wj\ (j=1,2)\;,
\label{heavy b}\\
  \tilde q_R &\to& q \tilde Z_i\ (i=1,2,3,4)\;. \label{heavy c}
\end{eqnarray}
Some decays go via loops (e.g.\ $\tilde g\to g \tilde Z_1)$; we have not
attempted an exhaustive listing here.
Figure~7 shows how gluino-to-heavy-gaugino branching fractions increase
with $m_{\tilde g}$ in a particular example
(with $m_{\tilde g} < m_{\tilde q}$)~\cite{bbkt}.

\begin{center}
{\small Fig.~7. Example of gluino decay branchings versus mass~\cite{bbkt}.}
\end{center}

   The heavier gauginos then decay too:
\begin{eqnarray}
  \tilde W_j &\to& Z \tilde W_k,\, W \tilde Z_i,\, H_i^0 \tilde W_k,\,
                 H^\pm \tilde Z_i,\,f\tilde f   \;, \\
  \tilde Z_i &\to& Z \tilde Z_k,\, W \tilde W_j,\, H_i^0 \tilde Z_k,\,
                 H^\pm \tilde W_k,\, f\tilde f' \;.   \label{Ztwid decay}
\end{eqnarray}
Here it is understood that final $W$ or $Z$ may be off-shell and
materialize as fermion-antifermion pairs; also $Z$ may be replaced
by $\gamma$.    To combine the complicated production and cascade
possibilities systematically, all these channels have been incorporated in
the ISAJET~7.0 Monte Carlo package called ISASUSY~\cite{bppt}.

   These multibranch cascade decays lead to higher-multiplicity final
states in which the LSPs $\tilde Z_1$  carry a much smaller share of the
available energy, so $\overlay/p_T$ is smaller and less distinctive,
making detection via $\overlay/p_T$ more difficult.
Remember that leptonic $W$ or $Z$ decays,
$\tau$ decays, plus semileptonic $b$ and $c$ decays, all give background events
with genuine $\overlay/p_T$; measurement uncertainties also contribute fake
$\overlay/p_T$ backgrounds.  Experimental bounds therefore become weaker when
we take account of cascade decays.  The CDF 90\% CL
limits on the $(m_{\tilde g}, m_{\tilde q})$ masses quoted above are reduced by
10--30~GeV when cascade decays are taken into account\cite{btw}.

Cascade decays also present new opportunities for SUSY detection.
 Same-sign dileptons (SSD) are a very interesting signal~\cite{bkp}, which
arises naturally from $\tilde g \tilde g$  and  $\tilde g \tilde q$  decays
because of the Majorana character of gluinos, with very
little background. Figure~8 gives an example of this signal.
Eqs.~(\ref{heavy a})--(\ref{Ztwid decay})  show how a heavy gluino or squark
can decay to a chargino
$\tilde W_j$ and hence, via a real or virtual $W$, to an isolated charged
lepton.  For such squark pair decays the two charginos --- and hence the two
leptons --- are constrained to have opposite signs, but if a gluino is
present it can decay equally into either sign of chargino and lepton
because it is a Majorana fermion.  Hence  $\tilde g \tilde g$  or
$\tilde g \tilde q$  systems can decay to isolated SSD plus jets plus
$\overlay/p_T$.  Cascade decays of $\tilde q\tilde q$ via the heavier
neutralinos $\tilde Z_i$ offer similar possibilites for SSD, since the $\tilde
Z_i$ are also Majorana fermions. Cross sections for the Tevatron are shown in
Fig.~9.

\begin{center}

{\small Fig.~8. Example of same-sign dilepton appearance in gluino-pair decay.}

\medskip

{\small Fig.~9. Same-sign dilepton signals at the Tevatron~\cite{baerev}.}
\end{center}

   Genuinely isolated SSD backgrounds come from the production of $WZ$ or
$Wt\bar t$ or $W^+W^+$ ({\it e.g.}\ $uu \to ddW^+W^+$ by gluon exchange), with
cross sections of order  $\alpha_2^2$   or   $\alpha_2\alpha_3^2$    or
$\alpha_2^2\alpha_3^2$  compared to $\alpha_3^2$ for gluino pair production,
so we expect to control them with suitable cuts.  Very large   $b \bar b$
production gives SSD via semileptonic $b$-decays plus $B$-$\bar B$ mixing, and
also via combined $b\to c\to s \,\ell^+ \nu$   and  $\bar b\to \bar c\,\ell^+
\nu$ decays, but both leptons are produced in jets and can be suppressed by
isolation cuts.  Also $t\bar t$ gives SSD via $t\to b\, \ell^+ \nu$ and  $\bar
t\to \bar b\to \bar c\, \ell^+ \nu$, but the latter lepton is non-isolated.
So SSD provide a promising SUSY signature.

   Gluino production rates at SSC/LHC are much higher than at the Tevatron.
At $\sqrt s=40$~TeV, the cross section is
\begin{equation}
\sigma(\tilde g \tilde g) = 10^4,\,700,\, 6\;\mbox{fb\quad for }
 m_{\tilde g} = 0.3,\,1,\,2\rm\; TeV\;.
\end{equation}
Many different SUSY signals have been evaluated, including
$\overlay/p_T + n\,$jets, $\overlay/p_T +{}$SSD, $\overlay/p_T + n\,$isolated
leptons, $\overlay/p_T + {}$one isolated lepton${}+ Z$, $\overlay/p_T + Z$,
$\overlay/p_T + Z + Z$.
SSC cross sections for some of these signals from  $\tilde g \tilde g$
production are shown versus $m_{\tilde g}$ in Fig.~10 (for two scenarios,
after various cuts);  the labels 3,4,5 refer to numbers of isolated
leptons~\cite{btw}.

\begin{center}
{\small Fig.~10. SSC cross sections for various SUSY signals, after
cuts~\cite{btw}.}
\end{center}

   Heavy gluinos can also decay copiously to $t$-quarks~\cite{btw,bps}:
\begin{equation}
\tilde g \to t \bar t\tilde Z_i , t \bar b \tilde W^-, b \bar t
\tilde W^+ \;.
\end{equation}
$t\to bW$  decay then leads to multiple $W$ production.  For example, for a
gluino of mass 1.5~TeV,  the  $\tilde g\to W,\, WW,\, WWZ,\, WWWW$ branching
fractions are typically of order 30\%, 30\%, 6\%, 6\%, respectively. For
$m_{\tilde g}\sim 1$~TeV the SUSY rate for $4W$ production can greatly
exceed the dominant SM $4t\to 4W$ mode, offering yet another signal for
SUSY~\cite{bps}.

To summarize this section:
\begin{enumerate}

\item Light SUSY particle searches are based
largely on $\overlay/p_T$ signals.  But for $m_{\tilde g},\,m_{\tilde q}
> 50$~GeV cascade decays become important; these cascades both weaken the
simple $\overlay/p_T$
 signals and provide new signals such as same-sign dileptons.

\item For even heavier squarks and gluinos, the cascade decays dominate
completely and provide further exotic (multi-$W,Z$ and multi-lepton)
signatures.

\item Gluinos and squarks in the expected mass range of Eq.~(\ref{m bound})
will not escape detection.
\end{enumerate}

\section{MSSM Higgs Phenomenology}

In minimal SUSY, two Higgs doublets $H_1$ and $H_2$ are needed to cancel
anomalies and at the same time give masses to both up- and down-type quarks.
Their vevs are $v_1=v\cos\beta$ and $v_2=v\sin\beta$. There are 5 physical
scalar states: $h$ and $H$ (neutral CP-even with $m_h<m_H$), $A$ (neutral
CP-odd) and $H^\pm$. At tree level the scalar masses and couplings and an
$h$-$H$ mixing angle $\alpha$ are all determined by two parameters,
conveniently chosen to be $m_A$ and $\tan\beta$. At tree level the masses obey
$m_h\le M_Z,m_A; m_H\ge M_Z,m_A; m_{H^\pm}\ge M_W,m_A$.

Radiative corrections are significant, however~\cite{rad}. The most important
new parameters entering here are the $t$ and $\tilde t$ masses; we neglect for
simplicity some other parameters related to squark mixing. One-loop corrections
give $h$ and $H$ mass shifts of order $\delta m^2\sim G_F\,m_t^4\ln(m_{\tilde
t}/m_t)$, arising from incomplete cancellation of $t$ and $\tilde t$ loops. The
$h$ and $H$ mass bounds get shifted up and for the typical case $m_t=150$~GeV,
$m_{\tilde t}=1$~TeV we get
\begin{equation}
m_h < 116\gev < m_H \,.
\end{equation}
There are also corrections to cubic $hAA,\,HAA,\,Hhh$ couplings, to $h$-$H$
mixing, and smaller corrections to the $H^\pm$ mass. Figure~11 illustrates the
dependence of $m_h$ and $m_H$ on $m_A$ and $\tan\beta$, for two different
values of $m_t$ (with $m_{\tilde t}=1$~TeV still). We shall assume $\tan\beta$
obeys the GUT constraints $1\le\tan\beta\le65$ of Eq.~(\ref{eff bound}).

At LEP\,I, the ALEPH, DELPHI, L3 and OPAL collaborations~\cite{LEP} have all
searched for the processes
\begin{equation}
e^+e^- \to Z \to Z^*h,Ah \,,
\end{equation}
with $Z^*\to\ell\ell,\nu\nu,jj$ and $h,A\to\tau\tau,jj$ decay modes. The $ZZh$
and $ZAh$ production vertices have complementary coupling-strength factors
$\sin(\beta-\alpha)$ and $\cos(\beta-\alpha)$, respectively, helping to give
good coverage. The absence of signals excludes regions of the $(m_A,\tan\beta)$
plane; Fig.~12 shows typical boundaries for various $m_t$ values, deduced from
ALEPH results~\cite{bcps,LEP}. These results imply lower bounds
\begin{equation}
m_h,m_A\agt20\mbox{--45~GeV (depending on}\tan\beta) \,.
\end{equation}
Null searches for $e^+e^-\to H^+H^-$ exclude a region with
$\tan\beta<1$~\cite{diaz}.

\begin{center}
\parbox{5.5in}{\small Fig.~11. Contours of $h$ and $H$ masses in the
$(m_A,\tan\beta)$ plane for (a)~$m_t=150$~GeV, (b)~$m_t=200$~GeV, with
$m_{\tilde t}=1$~TeV.}
\end{center}

\smallskip

\begin{center}
\smallskip
\parbox{5.5in}{\small Fig.~12. Limits from ALEPH searches for (a)~$Z\to Z^*h$
and (b)~$Z\to Ah$ at LEP\,I, for various $m_t$ values with $m_{\tilde
t}=1$~TeV~\cite{bcps,LEP}.}
\end{center}

LEP\,II will have higher energy and greater reach. Figure~13 shows approximate
discovery limits in the  $(m_A,\tan\beta)$ plane for various $m_t$ values,
based on projected searches for $e^+e^-\to ZH\to \ell\ell jj,\nu\nu jj,jjjj$
and for $e^+e^-\to (Zh,Ah)\to \tau\tau jj$, assuming energy $\sqrt s=200$~GeV
and luminosity ${\cal L}=500\rm\,pb^{-1}$. $H^\pm$ searches will not extend
this reach.

\begin{center}
\medskip
\parbox{5.5in}{\small Fig.~13. Projected limits for various LEP\,II searches,
assuming $\sqrt s=200$~GeV and ${\cal L}=500\rm\,pb^{-1}$~\cite{bcps}.}
\end{center}

Searches for neutral scalars at SSC and LHC will primarily be analogous to SM
Higgs searches:
\begin{enumerate}

\item untagged $\gamma\gamma$ signals from $pp\to(h,H,A)\to\gamma\gamma$ via
top quark loops (Fig.~14);

\item tagged $\gamma\gamma$ signals from $pp\to(h,H,A)\to\gamma\gamma$ plus
associated $t\bar t$ or $W$, permitting lepton tagging via $t\to W\to \ell\nu$
or $W\to\ell\nu$ decays (Fig.~15);

\item  four-lepton signals from $pp\to(h,H)\to ZZ$ or
$Z^*Z\to\ell^+\ell^+\ell^-\ell^-$ (Fig.~16).
\end{enumerate}

\noindent
Though qualitatively similar to SM signals, these will generally be smaller due
to the different coupling constants that depend on $\beta$ and $\alpha$.

\medskip

\begin{center}
{\small Fig.~14. Typical diagram for untagged Higgs${}\to\gamma\gamma$
signals.}


\medskip

{\small Fig.~15. Typical diagrams for lepton-tagged Higgs${}\to\gamma\gamma$
signals.}

\medskip

{\small Fig.~16. Typical diagrams for ``gold-plated'' four-lepton Higgs
signals.}
\end{center}

For charged Higgs scalars, the only copious hadroproduction source appears to
be top production with $t\to bH^+$ decay (that requires $m_{H^\pm}<m_t-m_b$).
The subsequent $H^+\to c\bar s,\nu\bar\tau$ decays are most readily detected in
the $\tau\nu$ channel (favored for $\tan\beta>1$), with $\tau\to\pi\nu$ decay
(Fig.~17).

\begin{center}

{\small Fig.~17. Typical diagram for $\tau$ signals from top decay via
charged-Higgs modes. }
\end{center}
\bigskip

\noindent
SM $t$-decays give equal probabilities for $e,\nu,\tau$ leptons via $t\to bW\to
b(e,\mu,\tau)\nu$, but the non-standard $t\to bH^+\to b\tau\nu$ leads to
characteristic excess of $\tau$. The strategy is to tag one top quark via
standard $t\to bW\to b\ell\nu$ decay and to study the $\tau/\ell$ ratio in the
associated top quark decay ($\ell=e$ or $\mu$).

Several groups have studied the detectability of these signals at SSC/LHC, and
they all reach broadly similar conclusions~\cite{bcps,baer,gunion,kunszt}.
Figure~18 shows typical limits of detectability for untagged and lepton-tagged
$\gamma\gamma$ signals at SSC, assuming luminosities ${\cal L}=20\,\rm fb^{-1}$
(two years of running) and $m_t=150$~GeV. Figure~19(a) shows a similar limit
for the $H\to4\ell$ search (no $h\to4\ell$ signal is detectable).  Figure~19(b)
shows typical limits for detecting the $t\to H^+\to{}$excess $\tau$ signal;
here the value of $m_t$ is critical, since only the range $m_{H^+}<m_t-m_b$ can
contribute at all. Putting all these discovery regions together with the LEP\,I
and LEP\,II regions, we see that very considerable coverage of the
$(m_A,\tan\beta)$ plane can be expected --- but there still remains a small
inaccessible region; see Fig.~20.
For $m_t=120$~GeV the inaccessible region is larger, for $m_t=200$~GeV it is
smaller.

\begin{center}

\parbox{5.75in}
{\small Fig.~18. Limits of detectability for $H,h,A$ $\gamma\gamma$ signals at
the SSC, for $\L=20\fb^{-1}$, (a)~without tagging, and (b)~with lepton
tag~\cite{bcps}.}

\medskip
\parbox{5.75in}
{\small Fig.~19. Detectability limit for (a) $H\to4\ell$ signals and (b)~$t\to
bH^+\to b\tau^+\nu$ signals
at the SSC for $\L=20\fb^{-1}$~\cite{bcps}.}
\end{center}

\begin{center}

\medskip
\parbox{5.5in}{\small Fig.~20. Combined LEP and SSC discovery regions for
$m_t=150$~GeV from Ref.~30; similar results are obtained by other
groups~\cite{baer,gunion,kunszt}.}

\end{center}

Figure 21 shows how many of the MSSM scalars $h,H,A,H^\pm$ would be detectable,
in various regions of the $(m_A,\tan\beta)$ plane. In many regions two or more
different scalars could be discovered, but for large $m_A$ only $h$ would be
discoverable; in the latter region, the $h$ couplings all reduce to SM
couplings, the other scalars become very heavy and approximately degenerate,
and the MSSM essentially behaves like the SM.

\begin{center}

{\small Fig.~21.  How many MSSM Higgs bosons may be discovered (from Ref.~30).}
\end{center}

An indirect constraint on the MSSM Higgs sector is provided by the CLEO bound
on $b\to s\gamma$ decays~\cite{cleo},
\begin{equation}
B(b\to s\gamma)<5.4\times10^{-4}\quad (95\%\ \rm C.L.)  \,.
\end{equation}
In the SM this decay proceeds via a $W$ loop process, but in models with more
than one Higgs doublet there are charged Higgs contributions too (Fig.~22). In
the MSSM both the $W$ and $H$ amplitudes have the same sign and the branching
fraction is directly related to $m_{H^+}$ and $\tan\beta$; hence the CLEO
result implies a lower bound on $m_{H^+}$ for given $\tan\beta$ [Fig.~23(a)].
It was recently pointed out~\cite{hewett,bbp} that this CLEO-based constraint
falls in a very interesting and sensitive region when translated to the
$(m_A,\tan\beta)$ plane; see Fig.~23(b). Taken at face value, it appears to
exclude a large part of the LEP\,II discovery region and even to exclude the
otherwise inaccessible region too.

\begin{center}
\medskip

{\small Fig.~22. $W$ and charged-Higgs loop diagrams contributing to $b\to
s\gamma$ decays.}

\medskip

\parbox{6in}{\small Fig.~23. (a) Lower bound on $m_{H^+}$ for given
$\tan\beta$, from $b\to s\gamma$ constraint. (b)~Comparison of $b\to s\gamma$
bound with other MSSM Higgs constraints in the $(m_A,\tan\beta)$ plane, for
$m_t=150$~GeV.
The regions excluded by the CLEO experimental bound are to the left of the
$b\to s\gamma$ curves; the curves shown are updated from Ref.~38. }

\end{center}

It is premature however to reach firm conclusions yet. The calculations of
Ref.~38 are based on the approximation of Ref.~39, but later work indicates
further small corrections~\cite{misiak}. More importantly, other SUSY loop
diagrams (especially chargino loops) can give additional contributions of
either sign, leading to potentially significant changes in the
amplitude~\cite{bertolini,barbieri}. However, as theoretical constraints on
SUSY particles become more extensive, and as the $B(b\to s\gamma)$ bound itself
becomes stronger, we may expect this approach to give a valuable constraint in
the MSSM Higgs phenomenology.

Finally, what could a future $e^+e^-$ collider do? We have seen that part of
the MSSM parameter space is inaccessible to LEP\,II. But a possible future
linear collider with higher energy and luminosity could in principle cover the
full parameter space. In is interesting to know what are the minimum $s$ and
$\L$ requirements for complete coverage, for given $m_t$. This question was
answered in Ref.~43, based on the conservative assumption that only the
channels $e^+e^-\to(Zh,Ah,ZH,AH)\to\tau\tau jj$ would be searched, with no
special tagging. The results are shown in Fig.~24. We have estimated that
including all $Z\to\ell\ell,\nu\nu,jj$ and $h,H,A\to bb,\tau\tau$ decay
channels plus efficient $b$-tagging could increase the net signal $S$ by a
factor~6 and the net background $B$ by a factor~4, approximately; this would
increase the statistical significance $S/\sqrt B$ by a factor~3 and hence
reduce the luminosity requirement by a factor~9 or so. In this optimistic
scenario, the luminosity axis in Fig.~24 would be rescaled downward by an order
of magnitude.

\begin{center}
\medskip

\parbox{5.5in}{\small Fig.~24. Conservative requirements for a ``no-lose'' MSSM
Higgs search at a future $e^+e^-$ collider. Curves of minimal $(\sqrt s,\L)$
pairings are shown for $m_t=120$, 150, 200~GeV; the no-lose region for
$m_t=150$~GeV is unshaded~\cite{nolose}.}
\end{center}

\bigskip

To summarize this Section:

\begin{enumerate}

\item The MSSM Higgs spectrum is richer but in some ways more elusive than the
SM case.

\item At least one light scalar is expected.

\item As $m_A\to\infty$ this light scalar behaves like the SM scalar and the
others become heavy.

\item LEP\,I, LEP\,II and SSC/LHC will give extensive but not quite complete
coverage of the MSSM parameter space.

\item For some parameter regions, several different scalars are detectable, but
generally one or more remain undetectable.

\item The $b\to s\gamma$ bound has the potential to exclude large areas of
parameter space (possibly including the inaccessible region) but is presently
subject to some uncertainty.

\item A higher-energy $e^+e^-$ collider could cover the whole MSSM parameter
space, discovering at least the lightest scalar $h$.

\end{enumerate}

\section*{Acknowledgements}

We thank H.~Baer, M.~Berger, and P.~Ohmann for valuable contributions to the
contents of this review. This work was supported in part by the University of
Wisconsin Research Committee with funds granted by
the Wisconsin Alumni Research Foundation, in part by the U.S.~Department of
Energy under contract no.~DE-AC02-76ER00881, and in part by the Texas National
Laboratory Research Commission under grant no.~RGFY9273.

\end{document}